# Physics of Top Quark at the Tevatron

C.–P. Yuan

*Department of Physics and Astronomy*
*Michigan State University*
*East Lansing, MI 48824, USA*

**Abstract.** We discuss the physics of the top quark at the Fermilab Tevatron. By the year 2000, many properties of the top quark can be measured at the Tevatron.

## 1 Discovery of the Top Quark

The discovery of the top quark is one of the most important achievements at the Fermilab Tevatron which is currently the only collider that produces the top quark on its mass-shell. The mass of the top quark has been measured to be $m_t = 176 \pm 8$ (stat.)$\pm 10$(sys.) GeV by the CDF group and $m_t = 199^{+19}_{-21}$ (stat.)$\pm 22$(sys.) GeV by the DØ group, through the detection of $t\bar{t}$ events. The standard $t\bar{t}$ event selection is based on the expected Standard Model (SM) decay chain $t\bar{t} \to (W^+b)(W^-\bar{b})$ and the subsequent decays of the $W$'s into fermion pairs. At least one $W$ is tagged in the mode $W \to \ell + \nu$ by requiring an isolated high $p_T$ (transverse momentum) lepton ($\ell = e$ or $\mu$) and large $\slashed{E}_T$ (missing transverse energy). In the "dilepton" analysis the leptonic decay of the other $W$ is identified with a loose lepton selection; this mode has small backgrounds, but also a small branching ratio of just $4/81 \simeq 5\%$. In the case of the "lepton + jets" mode, the second $W$ decays to quark pairs, giving larger branching ratio of $24/81 \simeq 30\%$. The final state of $(\ell\nu b)(jj\bar{b})$ is separated from the primary background, $W$+jets, by requiring a large multiplicity of high $p_T$ jets and also evidence of a $b$-jet, using either secondary vertex (silicon detector) or soft lepton ($b \to c\ell\nu X$) identification. For more detailed discussions on the event selection and the detector configuration which determines the acceptance and the detection efficiency of the events, we refer the readers to Ref. [1].

Assuming the SM decay mode of the top quark, the cross section $\sigma_{t\bar{t}}$ for the QCD production processes $q\bar{q}, gg \to t\bar{t}$ was measured to be $7.6^{+2.4}_{-2.0}$ pb by CDF and $6.4 \pm 2.2$ pb by DØ. For comparison, the SM result for $m_t = 175$ GeV at $\sqrt{s} = 1.8$ TeV is $\sigma_{t\bar{t}} = 5.52^{+0.07}_{-0.45}$ pb quoted from Ref. [2] in which the effects of multiple soft-gluon emissions have been properly resummed. Since the measurement of the cross section obtained from the "counting" experiments (counting the observed total $t\bar{t}$ event numbers in various decay modes) and the measurement of the mass of the top quark (obtained from reconstructing the invariant mass of the top quark) are not strongly correlated, one can combine these results to find the best fitted values for $m_t$ and $\sigma_{t\bar{t}}$ [3].

It is important to remember that these experiments have optimized their search for the process $p\bar{p} \to t\bar{t}X \to bW^+\bar{b}W^-X$, so they actually report the product of the top quark production cross section $\sigma_{t\bar{t}}$ and the branching ratio squared $b^2$, where $b = \mathrm{BR}(t \to bW)$. Furthermore, the reported production cross sections are a function of the assumed top mass value used in the analysis. In Ref. [4], we updated the above fit by including the measured values from CDF and DØ for $m_t$ and $\sigma_{t\bar{t}} \times b^2$. Finding the minimum value of $\chi^2$ yields $m_t = 168.6^{+3.0}_{-3.0}$ GeV, $\sigma_{t\bar{t}} = 7.09^{+.68}_{-.62}$ pb and $b = 1.00^{+.00}_{-.13}$.[1] At the 95% confidence level (C.L.) [2], $b = .74$. This number, then, gives us an upper limit on $\mathrm{BR}(t \to X)$, where $X \neq bW$. From the results of the fit described above, we conclude that $\mathrm{BR}(t \to X)$ for $X \neq bW$ has to be less than $\sim 25\%$. This result agrees with a measurement of $b = .87^{+.18}_{-.32}$ by CDF based on single- and double-$b$-tagged events [5].

Obviously, measuring $m_t$ and $\sigma_{t\bar{t}}$ is not all the Tevatron can/will do. In this paper, we will also discuss the single-top production cross section, some features of the $t$-$b$-$W$ vertex, decay branching ratios, partial decay width and lifetime of the top quark, and some non-SM decay modes and exotic production mechanisms.

## 2 Is the Top Quark just another heavy quark?

Because the top quark is heavy and its mass is of the order of the electroweak symmetry breaking (EWSB) scale $v = (\sqrt{2}G_F)^{-1/2} = 246$ GeV, its study might provide clues to the generation of the fermion masses which could be closely related to the EWSB. Furthermore, effects of new physics originating from the EWSB would be more apparent in the top quark sector than any other light sector of the electroweak theory. Hence, the top quark system not only serves as a stage for testing the SM but also provides a window to new physics beyond the SM.

A few examples are discussed in Ref. [6] to illustrate that different models of EWSB mechanism will induce different interactions among the top quark and the $W$- and $Z$-bosons. These interactions may strongly modify the production and/or the decay of the top quark. In some models (e.g., the TopColor model [7]) observable flavor-changing neutral current (FCNC) processes (e.g., $t \to cZ$, $cg$, $c\gamma$ ...) can be mediated by new underlying dynamics, and some new resonances can strongly couple to $t\bar{t}$ (e.g., a degenerate, massive color octet of "colorons" and a singlet heavy $Z'$) or $t\bar{b}$ (e.g., a triplet of "top-pions") system. With all these new effects possibly appearing in the top quark system, we conclude that the top quark is likely to be not just another heavy quark.

---
[1] The theoretical prediction of $\sigma_{t\bar{t}}$ is 6.83 pb for $m_t = 168.6$ GeV.
[2] We varied the parameter until the $\chi^2$ value increased from its minimal value by $(1.96)^2$ units.

Is it a SM top quark? What do we know about the interactions of the top quark? Can we learn about them from the radiative effects on the precision LEP/SLC data (physics at the $Z$-pole)? A few analyses for studying the top quark couplings to the gauge bosons show that current low energy data still allow rooms for new physics [6, 8], and that only the direct measurements (with on-shell top quark produced) of these couplings can be conclusive.

## 3 Top Quark Physics for the Tevatron at Run-II and Beyond

To determine how well an observable can be measured at the Tevatron in the Main Injector Era (Run-II and beyond), we need to set up a reference for top quark event rates. For this purpose, we consider a $\bar{p}p$ collider with $\sqrt{S} = 2$ TeV and an integrated luminosity of 2 fb$^{-1}$ (or, 1 fb$^{-1}$ for each experimental group). In the SM, for a 175 GeV top quark, there will be about $1.4 \times 10^4$ $t\bar{t}$ pairs and $5 \times 10^3$ single-$t$ or single-$\bar{t}$ events produced. After taking into account the $b$-tagging efficiency and the detection efficiency [1], there are about 1000 single-$b$-tagged $t\bar{t}$ pairs in the $\ell +$ jets sample (among those 600 are also double-$b$-tagged), 100 in the dilepton sample, and 250 single-$t$ or single-$\bar{t}$ events (in the $\ell +$ jets sample) available for testing various properties of the top quark. In the following, we discuss the relevant observables and show that with a 2 fb$^{-1}$ luminosity, many *first* measurements can already be done to a good accuracy at the Tevatron. With a $(10-100)$ fb$^{-1}$ integrated luminosity (beyond Run-II), many further improvements are expected.

### 3.1  $t\bar{t}$ production rate $\sigma_{t\bar{t}}$

At the Tevatron, the dominant $t\bar{t}$ pair production mechanism is $q\bar{q} \to t\bar{t}$ not $gg \to t\bar{t}$, the former contributes about 90% of the rate because the quark luminosities are larger than the gluon luminosities for large x (i.e. for producing heavy top quarks). To test QCD, we need an accurate measurement for $\sigma_{t\bar{t}}$ which is experimentally limited by the systematic uncertainties. It is concluded in Ref. [1] that the experimental error in $\sigma_{t\bar{t}}$ is $\delta\sigma_{t\bar{t}} \simeq 10\%$ which is about the same as the theoretical error for calculating $\sigma_{t\bar{t}}$ [2]. Assuming the decay branching ratios of the top quarks are as described by the SM, $\sigma_{t\bar{t}}$ can also be measured by counting the production rate of dilepton events. Since the statistical error for the dilepton sample of $t\bar{t}$ events is about $1/\sqrt{100} = 10\%$, we expect $\delta\sigma_{t\bar{t}} \simeq 10\%$, as implied from this measurement.

### 3.2  Mass of the top quark $m_t$

With enough $t\bar{t}$ pair events, the accuracy in measuring $m_t$ will be determined by the systematic uncertainty which is dominated by the error in measuring the jet energy scale due to the imperfections in the calorimetry and the effects of initial and final state gluon radiations. The determination on the jet energy scale can be greatly improved by studying the $Z+1$ jet and $\gamma+1$ jet events [1]. It is expected that $m_t$ can be measured to within a couple percent. So, the uncertainty in measuring $m_t$ is $\delta m_t \simeq (2-4)$ GeV. With this accuracy on $m_t$, it becomes possible to get a useful constraint (within $\sim 100$ GeV) on the mass of the SM Higgs boson.

### 3.3  Distributions of invariant mass $M_{t\bar{t}}$ and transverse momentum $p_T(t)$

If a heavy new resonance $(V)$ can be produced in a $\bar{p}p$ collision and can strongly couple to $t\bar{t}$ [9], then the observed distributions of $M_{t\bar{t}}$ and $p_T(t)$ will be different from the SM predictions. (The event rates can either increase or decrease.) By carefully comparing these distributions with the predictions of a given theory, one can then either approve or exclude that theory. Since $M_{t\bar{t}}$ can be reconstructed on an event-by-event basis by requiring that there are two top quarks observed in the event, the shape and the magnitude of this distribution can therefore set a model-independent limit on possible new physics coupled to $t\bar{t}$ pairs. Demanding a resonance to be observed at the $5\sigma$ level (i.e. $S/\sqrt{B} \gtrsim 5$) over the $t\bar{t}$ continuum in the $M_{t\bar{t}}$ spectrum, one can set the minimum bound on $\sigma(\bar{p}p \to V) * (V \to t\bar{t})$ to about $(0.4-0.8)$ pb and $(0.1-0.2)$ pb for $M_V$ equal to 500 GeV and 800 GeV, respectively [1].

### 3.4  Top quark decays and FCNC decay modes of top

Because the top quark is heavy, it will decay via weak interaction before it feels non-perturbative strong interaction. This is the first opportunity we have for studying the properties of a bare quark. In the SM, the total decay width of a SM top quark is $\Gamma_t \simeq 1.6\,\text{GeV}(m_t/180\,\text{GeV})^3$, and the branching ratio of the weak two body decay $t \to bW^+$ is about one hundred percent. In this decay mode the top quark will analyze its own polarization [10]. If $t$ is found dominantly decaying to $bW^+$, the second top $(\bar{t})$ in each $t\bar{t}$ event should be carefully studied as a window for small non-SM decay modes of top quarks. For instance, in the Minimum Supersymmetric Standard Model (MSSM), $t$ can decay into $bH^+$, $t \to \tilde{t}_1 \tilde{\chi}_1^0$, or $R$-parity violating channels, etc., and their branching ratios depend on the detailed parameters of the model. With enough $t\bar{t}$ events, one can tag one $t$ and study the detailed properties of another $t$ for each $t\bar{t}$ event. We call this $t$-tagging.

In the SM, the branching ratios for the FCNC decay modes were found to be too small to be detected, for instance, $Br(t \to cH) \sim 10^{-7}$, $Br(t \to cg) \sim 10^{-10}$, $Br(t \to cZ) \sim 10^{-12}$, and $Br(t \to c\gamma) \sim 10^{-12}$. However, in some models, the FCNC decay modes of the top quark can be observable [11]. Consider the MSSM with light chargino and top-squark. One of the $t$ in the $t\bar{t}$ event decays to $\tilde{t}_1 \tilde{\chi}_1^0$, and $\tilde{t}_1$ subsequently decays to $c\tilde{\chi}_1^0$. The signature of this event is $W + 2\,\text{jet} + \not{E}_T$ which is not included in the counting experiments that only count events with $W + \geq 3\,\text{jets}$. A careful study on this signature can approve the MSSM or set limit on the MSSM parameters [4]. Other studies [1] show that a $2\,\text{fb}^{-1}$ luminosity can be sensitive to $\text{BR}(t \to cZ) \sim 2\%$ (from $3\ell + 2\,\text{jets}$ or $2\ell + 4\,\text{jets}$ sample) and $\text{BR}(t \to c\gamma) \sim 0.3\%$ (from $\ell + \gamma + 2\,\text{jets}$ or $\gamma + 4\,\text{jets}$ sample).

### 3.5 *Ratios of branching ratios: $\mathcal{R}_\ell$ and $\mathcal{R}_b$*

Define $\mathcal{R}_\ell$ to be the ratio of the $t\bar{t}$ cross sections measured using dilepton events to that measured using $\ell + \text{jets}$ events. If $\mathcal{R}_\ell$ differs from the SM prediction, then it will imply new physics that will allow $t$ to decay without a $W$ boson in the final state, as in a charged Higgs decay ($t \to bH^+$ [12]), or a top-squark decay ($t \to \tilde{t}_1 \tilde{\chi}_1^0$). Hence, $\mathcal{R}_\ell$ measures $\text{BR}(t \to bW)$. With $2\,\text{fb}^{-1}$, the error on $\text{BR}(t \to bW)$ is about 10% [1]. Another useful ratio $\mathcal{R}_b$ is that of the $t\bar{t}$ cross section measured by using double-$b$-tagged events to the one measured by using single-$b$-tagged events. This determines the upper limit on the branching ratio of $t \to X$ where $X$ does not contain any $b$-quark. This method can be applied to both $\ell + \text{jets}$ and dilepton samples, from a known $b$-tagging efficiency. With $2\,\text{fb}^{-1}$, this upper limit can be set to about $(3-5)\%$ [1]. This result can be interpreted as the error on measuring $\text{BR}(t \to bW)$ if a $W$ boson is confirmed in $t$ decay.

With a large $t\bar{t}$ event rate, one can use $\mathcal{R}_\ell$ and $\mathcal{R}_b$ to select events in which the top quark decays to a $W$-boson and a $b$-quark, and then redo the study as done in Ref. [4] for obtaining the best fit on $\sigma_{t\bar{t}}$, $m_t$ and $\text{BR}(t \to bW)$. Generally speaking, a small $\text{BR}(t \to X)$ is better measured from direct search of the rare decay mode than from the measurements of $\mathcal{R}_\ell$ and $\mathcal{R}_b$.

### 3.6 *Partial decay width $\Gamma(t \to bW)$ and the lifetime of top*

The total decay width $\Gamma_t$ of a SM top quark cannot be measured from the invariant mass of $t$ reconstructed in the $t\bar{t}$ events [13]. An elegant way to determine the lifetime (the inverse of the total decay width $\Gamma_t$) of the top quark is to measure the partial decay width $\Gamma(t \to bW)$ and the branching ratio $\mathrm{BR}(t \to bW)$. This is because $\Gamma_t = \Gamma(t \to bW)/\mathrm{BR}(t \to bW)$.

As discussed above, $\mathrm{BR}(t \to bW)$ can be determined from measuring $\mathcal{R}_\ell$ and $\mathcal{R}_b$. The width $\Gamma(t \to bW^+)$ can be measured by counting the production rate of top quarks from the $W$-$b$ fusion process which is *equivalent* to the $W$-gluon fusion process by properly treating the bottom quark and the $W$ boson as partons inside the hadron [13]. Consider the $q'b \to qt$ process. It can be viewed as the production of an on-shell $W$-boson which then rescatters with the $b$-quark to produce the top quark. This is known as the "effective-$W$ approximation". Although some assumptions about the dynamics of the hard scattering (for approximating an off-shell $W$-boson by an on-shell $W$-boson) need to be made for applying such an approximation, the kinematics of this factorization is exactly the same as that in the deep-inelastic scattering processes. The analytic expression for the flux ($f_\lambda(x)$) of the incoming $W_\lambda$-boson ($\lambda = 0, +, -$ for longitudinal, right-handed, or left-handed polarization) to rescatter with the $b$-quark can be found in Ref. [14]. The constituent cross section of $ub \to dt$ is given by

$$\hat{\sigma}(ub \to dt) = \sum_{\lambda=0,+,-} f_\lambda\left(x = \frac{m_t^2}{\hat{s}}\right) \left[\frac{16\pi^2 m_t^3}{\hat{s}(m_t^2 - M_W^2)^2}\right] \Gamma(t \to bW_\lambda^+),$$

where $M_W$ is the mass of $W^+$-boson and $\sqrt{\hat{s}}$ is the invariant mass of the hard part process. Hence, the production rate of single-top event from the $W$-gluon fusion process measures the partial decay width of the top quark $\Gamma(t \to bW^+)$.

At the Run-II of the Tevatron (with an integrated luminosity of $2\,\mathrm{fb}^{-1}$), we expect that the lifetime of the top quark will be known to about $20\% \sim 30\%$. Here, we have assumed that the branching ratio for $t \to Wb$ can be measured to about $10\%$ and the cross section for $W$-gluon fusion process is known to about $15\% \sim 20\%$ [13]. The total cross section for the $W$-gluon process was calculated by applying the QCD subtraction procedure described in Ref. [15] to properly resum all the large logs of the form $[\alpha_s \ln(Q/m_b)]^n$ from $n$-fold collinear gluon emission.

### 3.7  *Form factors of t-b-W , $V_{tb}$ and $m_t$*

The most general operators for the $t$-$b$-$W$ coupling are [10]:

$$\frac{g}{\sqrt{2}} \left[ W_\mu^- \bar{b}\gamma^\mu(f_1^L P_- + f_1^R P_+)t - \frac{1}{M_W}\partial_\nu W_\mu^- \bar{b}\sigma^{\mu\nu}(f_2^L P_- + f_2^R P_+)t \right]$$
$$+ \frac{g}{\sqrt{2}} \left[ W_\mu^+ \bar{t}\gamma^\mu(f_1^{L*} P_- + f_1^{R*} P_+)b - \frac{1}{M_W}\partial_\nu W_\mu^+ \bar{t}\sigma^{\mu\nu}(f_2^{R*} P_- + f_2^{L*} P_+)b \right]$$
$$+ \partial^\mu W_\mu^- \bar{b}(f_3^L P_- + f_3^R P_+)t + \partial^\mu W_\mu^+ \bar{t}(f_3^{R*} P_- + f_3^{L*} P_+)b,$$

where $P_\pm = \frac{1}{2}(1 \pm \gamma_5)$, $i\sigma^{\mu\nu} = -\frac{1}{2}[\gamma^\mu, \gamma^\nu]$ and the superscript $*$ denotes the complex conjugate. In the SM, the only nonvanishing form factor at tree level is $f_1^L = 1$. These form factors will have different values if new physics exists. The typical energy transfer scale for the $t$-$b$-$W$ coupling in the decay of the top quark (in either the $t\bar{t}$ pairs or the single-top events) is of the order $M_W$. In the $W$-gluon fusion process, the $W$-boson is not on-shell, but under the "effective-$W$ approximation" this scale is again of order $M_W$. (This is why the production rate of this process can be related to the decay width $\Gamma(t \to bW^+)$.) In the Drell-Yan type single-top production process, $qq' \to W^* \to t\bar{b}$, the typical energy transfer scale is of order the invariant mass of the $t\bar{b}$ pair which is larger than $m_t$. Hence, for instance, if there is a new heavy resonance that couples to the $t\bar{b}$ pair, it will be likely to enhance the production rate of single-top events at the Tevatron from the $W^*$ process. However, in this case, neither the decay width of the top quark nor the single-top production rate from the $W$-gluon fusion process will be largely modified. Therefore, to fully explore the form factors for the interaction of $t$, $b$ and $W$, all these processes have to be detected.

Using the invariant mass $(m_{b\ell})$ of $b$ and $\ell$, one can determine the polarization of the $W$ boson which depends on the values of these form factors. A study [15] shows that if $f_1^L \approx 1$ and $f_1^R$ and $f_2^{L,R}$ are almost zero (SM values), then the errors $\delta f_1^L \sim (2-3)\%$ and $\delta f_1^R \sim 0.2$. The measurement of $f_1^L$ can be interpreted as the measurement of the CKM element $V_{tb}$ which is therefore known to within 3%. (We note that this result is more accurate than that obtained from measuring the $qq' \to W^* \to t\bar{b}$ production rate which yields a 10% error in measuring $V_{tb}$ [1].) Furthermore, the fraction $(F_L)$ of longitudinal $W$'s from top decays is equal to $\frac{1}{2}\frac{m_t^2}{M_W^2}$, and is independent of $f_1^{L,R}$, it is therefore a good tool for measuring $m_t$ [10]. (The quadratic dependence of $F_L$ on $m_t$ helps in $\delta m_t$ by a factor of 2.) A 2% accuracy in determining $F_L$ yields a $\sim 2\,\text{GeV}$ error in measuring $m_t$.

### 3.8 *Exotic production mechanism and testing CP violation in top*

If new physics modifies strongly the $t$-$c$-$g$ coupling, then the production rate of $t$-$c$ pairs from $q\bar{q} \to g \to t\bar{c}$ can be largely enhanced. By measuring this production rate at high $M_{tc}$, one can set the minimum energy scale $\Lambda_{tcg}$ at which new physics must set in. A study in Ref. [16] showed that $\Lambda_{tcg} \gtrsim 4\,\text{TeV}$ if no signal is found at a $3\sigma$ level. For this value of $\Lambda_{tcg}$, the branching ratio $t \to cg$ should be less than about 10%.

Besides all the potential physics discussed above, the Tevatron, as a $\bar{p}p$ collider, is unique for being able to test CP violation by measuring the production rates of single-top events. A nonvanishing asymmetry $(\mathcal{A}_t)$ in the inclusive production rates of the single-$t$ events and the single-$\bar{t}$ events signals CP violation [17]. With $2\,\mathrm{fb}^{-1}$, it is possible to observe this asymmetry for $\mathcal{A}_t \gtrsim 20\%$ [17].

This work is supported in part by the NSF under grant no. PHY-9309902. We thank our colleagues who contributed to the studies in Ref. [1].

# References

1. Dan Amidei and Chip Brock, "Report of the *TeV*2000 Study Group on Future ElectroWeak Physics at the Tevatron", 1995; and references therein.
2. E.L. Berger and H. Contopanagos, ANL-HEP-PR-95-31, May 1995.
3. D. Soper, talk at the QCD session of the XXX Rencontre de Moriond, Les Arcs, France, March 1995.
4. S. Mrenna and C.–P. Yuan, ANL-HEP-PR-95-66, 1995.
5. The CDF Collaboration and J. Incandela, "CDF Top Quark Production and Mass", FERMILAB–CONF–95/237–E, July 1995.
6. Ehab Malkawi and C.–P. Yuan, *Phys. Rev.* **D50**, 4462 (1994).
7. C.T. Hill, *Phys. Lett.* **B345**, 483 (1995).
8. J. Feliciano, F. Larios, R. Martinez and M.A. Perez, CINVESTAV-FIS-09-95, 1995; 
   T. Han, R.D. Peccei and X. Zhang, Fermilab-pub-95/160-T, 1995.
9. C.T. Hill and S. Parke, *Phys. Rev.* **D49**, 4454 (1994); 
   E. Eichten and K. Lane, *Phys. Lett.* **B327**, 129 (1994).
10. G.L. Kane, G.A. Ladinsky and C.–P. Yuan, *Phys. Rev.* **D45**, 124 (1992).
11. For instance, J.L. Diaz-Cruz, R. Martinez, M.A. Perez, and A. Rosado, *Phys. Rev.* **D41**, 891 (1990); 
    J.L. Diaz-Cruz and G. Lopez Castro, *Phys. Lett.* **B301**, 405 (1993).
12. J. Guasch, R. Jimenez and J. Sola, UAB-FT-370, July 1995.
13. D.O. Carlson and C.–P. Yuan, MSUHEP-50823, August 1995.
14. P.W. Johnson, F.I. Olness and Wu-Ki Tung, *Phys. Rev.* **D36**, 291 (1987); and the references therein.
15. D.O. Carlson, Ph.D. thesis, Michigan State University, MSUHEP-050727, August 1995; and the references therein.
16. E. Malkawi and T. Tait, MSUHEP-51116, 1995.
17. C.–P. Yuan, *Mod. Phys. Lett.* **A10**, 627 (1995).